\documentclass[12pt,dvips]{article}

\usepackage{rotating}
\usepackage{epsfig}
\usepackage{color}
\usepackage{hhline}
\usepackage{graphicx}
\usepackage{url}
\usepackage{amsmath,amssymb,bbold}

\begin{document}

\vspace{2cm}

\begin{center}
{\large \bf \boldmath{$Z$}-boson polarization as a
        model-discrimination analyzer}\\[1.5cm]
{Seong Youl Choi$^{1}$\footnote{e-mail: choisy7@gmail.com} } \\[0.5cm]
{\it  $^1$Department of Physics and RIPC, Chonbuk National University,
          Jeonju 54896, Korea}
\end{center}

\vskip 2.cm

\begin{abstract}
\noindent
Determining the spin of any new particle is critical in identifying
the true theory among various extensions of the Standard Model (SM).
The degree of $Z$-boson polarization in any two-body decay process
$A\to B Z$ is sensitive to the spin assignments of two new particles
$A$ and $B$. Considering all possible spin-0, 1/2 and 1 combinations
in a renormalizable field theory, we demonstrate that $Z$-boson polarization can indeed play a role of a decisive and universal
analyzer in distinguishing the different spin assignments.\\[2mm]
\end{abstract}

\newpage


\section{Introduction}

As the electroweak (EW) scale ($v=246$ GeV) and beyond are being explored
at the Large Hadron Collider (LHC), it is highly anticipated that
the true theory for the origin and stability of the
EW scale~\cite{Weinberg:1976,Weinberg:1979bn,Susskind:1979,tHooft:1980}
will be revealed.

One generic prediction in most of new models is the presence of new
particles partnered with some or all of the SM particles.
For instance, every SM particle in low-energy supersymmetry
(SUSY)~\cite{Nilles,Haber_Kane,Chung:2003fi,Drees,Binetruy,Wess_Bagger}
has a heavier partner whose spin differs
by 1/2. Alternatively, in universal extra dimension (UED)
models~\cite{Appelquist:2000nn}, each SM particle is paired with a tower
of Kaluza-Klein (KK) excitations with identical spin.
Thus, model-independent spin measurements are crucial in
discriminating among new scenarios.

Since the rest frame of the decaying particle is hardly reconstructible,
the direct spin measurement at the LHC are performed through
the Lorentz-invariant masses in sufficiently long decay
chains~\cite{Wang:2006hk,Smillie:2005ar,Wang_Yavin}.
Such spin-determination methods, however, rely heavily on the final
state spins and the chiral structure of couplings~\cite{Wang:2008sw}.

In this paper we analyze the two-body decay of a new heavy state $A$
into a new lighter state $B$ and an on-shell neutral $Z$-boson as
\begin{eqnarray}
A\to B\, +\, Z\, \to\, B\, +\, \ell^+\ell^-\,,
\label{eq:two_body_decay}
\end{eqnarray}
where the $Z$-boson polarization can be measured through the lepton
angular distributions in the decay $Z\to\ell^+\ell^-$ with respect
to the $Z$ flight direction, reconstructed with great precision.
If its branching fraction is sizable, the leptonic decay
(\ref{eq:two_body_decay}) is highly expected to be a promising tool
not only for diagnosing the properties of the
new particles~\cite{Choi:2003fs,Kim:2007zzm} but also for determining
the $A$ and $B$ spins together.

\begin{table}[htb]
\caption{\it Possible spin assignments to the states $A$ and $B$ in
         the decay $A\to B\, Z$ and typical processes in SUSY, UED
         and LH models, respectively, with the same event topology
         as the process (\ref{eq:two_body_decay}), when kinematically
         accessible. The labels F, S and V in the left-most
         column denote a spin-1/2 fermion, a spin-0 scalar and
         a spin-1 vector boson, respectively. }
\vskip 0.5cm
\centering
\begin{tabular}{|c||c|c|c|}
\hline
{} & { } & { } & { } \\
AB & SUSY & UED & LH \\
{} & { } & { } & { } \\ \hline\hline
{} & { } & { } & { } \\
FF$^{ }$ & $\tilde{\chi}^0_i\to\tilde{\chi}^0_j Z$
   & $f_{1h}\to f_{1l} Z$
   & $T\to tZ$ \\
{} & $\tilde{\chi}^\pm_i\to\tilde{\chi}^\pm_j Z$
   &  --
   &  --
   \\
{} & { } & { } & { } \\ \hline
{} & { } & { } & { } \\
SS$^{ }$ & $\tilde{f}^{ }_2\to\tilde{f}^{ }_1 Z$
   & --
   & $\phi^P\to HZ$ \\
{} & $A\to h Z$
   &  --
   &  --
   \\
{} & { } & { } & { } \\ \hline
{} & { } & { } & { } \\
VS$^{ }$ & --
   & $W^\pm_1\to H^\pm_1 Z$
   & $Z_H \to HZ $ \\
{} &  --
   &  --
   & $\gamma_H\to HZ$ \\
{} & { } & { } & { } \\ \hline
{} & { } & { } & { } \\
VV$^{ }$ &  --
   &  --
   & $ W^\pm_H \to W^\pm Z$ \\
{} & { } & { } & { } \\
\hline
\end{tabular}
\label{tab:typical_process}
\end{table}

If kinematically allowed, several two-body decays like the event topology
as the process (\ref{eq:two_body_decay}) can occur in each extension of
the SM like SUSY, UED and little Higgs (LH) models
(See Table~\ref{tab:typical_process}).

\section{$Z$ Polarization in the Rest Frame}

Before describing the two-body decays $A\to B\, Z$ in detail, we briefly
summarize how to reconstruct the $Z$-polarization through the lepton-angle
distributions of the leptonic $Z$-boson decays $Z\to\ell^+\ell^-$,
in particular, with $\ell=e$ and $\mu$. In the rest frame of the decaying
$Z$ boson reconstructed with great precision by measuring the lepton
momenta, the normalized $\ell^-$ polar-angle distributions with respect to
the $Z$ polarization axis defined to be the $Z$ flight direction in the
laboratory frame are given by
\begin{eqnarray}
   \frac{ d D_\pm }{ d\cos\theta_\ell }
&=& \frac{3}{8} \left(1+\cos^2\theta_\ell
                     \pm 2\xi_\ell \cos\theta_\ell\right)\,, \\
   \frac{ d D_0}{d\cos\theta_\ell}
&=& \frac{3}{4} \left(1-\cos^2\theta_\ell\right)\,,
\label{eq:lepton_angle_distribution}
\end{eqnarray}
for transversely-polarized (helicity = $\pm 1$) and longitudinally-polarized
(helicity = $0$) $Z$ bosons, respectively, with
$\xi_\ell = 2 v_\ell a_\ell/(v^2_\ell+a^2_\ell)$.
(Although in principle the $\ell^-$ angular distributions for $Z$ bosons
with positive and negative helicity can be studied separately, but they
are not easy to distinguish practically because of the very small LR
asymmetry factor $\xi_\ell \simeq -0.147$ used for their distinction
with $v_\ell =s^2_W-1/4\simeq -0.02$ and $a_\ell =1/4$.)
Note that the polar-angle distribution can be determined even without
knowing the full kinematics of the decay $A\to B\, Z$.

The general analysis of the $Z$-boson polarization developed in
the two-body decay $A\to B\, Z$ is most transparent if performed in
the helicity formalism~\cite{Jacob_Wick}. In the $A$ rest frame the
decay helicity amplitude
can be decomposed in terms of the decay polar and azimuthal angles for
the momentum direction of the $Z$ boson produced in the $A$ rest frame as
\begin{eqnarray}
 {\cal D}_{\sigma_A: \lambda\sigma_B}(\Theta,  \Phi)
=
 {\cal T}_{\lambda, \sigma_B}\,
 d^{j_A}_{\sigma_A,\lambda-\sigma_B}(\Theta)\,
 e^{i(\sigma_A-\lambda+\sigma_B)\Phi}\quad
\mbox{with}\quad
|\lambda-\sigma_B|\leq j_A\,,
\end{eqnarray}
where $j_A,\sigma_A$ is the spin and helicity of the particle $A$ and
$\sigma_B, \lambda$ are the helicities of the particle $B$ and $Z$ boson,
respectively.
(For the sake of discussion the $Z$ momentum direction will be referred
to as the production axis in the following.) Because of rotational
invariance, the reduced matrix elements ${\cal T}_{\lambda,\sigma_B}$,
which contains all the dynamical information on the decay process, is independent of the $A$ helicity.

After integrating the absolute square of the amplitude over the angles
and summing it over the $A$ and $B$ helicities, we can obtain
the longitudinal and transverse polarizations, and the LR polarization
asymmetry of the produced $Z$ boson in the $A$ rest frame,
\begin{eqnarray}
P^A_L
 &=& \frac{\sum_{\sigma_B} |{\cal T}_{0,\sigma_B}|^2}{
    \sum_\lambda\sum_{\sigma_B}
          |{\cal T}_{\lambda,\sigma_B}|^2}\,, \\
P^A_T
 &=& 1- P^A_L\,, \\
A^A_\pm
 &=& \frac{\sum_{\sigma_B}
    \left(|{\cal T}_{+1,\sigma_B}|^2
         -|{\cal T}_{-1,\sigma_B}|^2 \right)}
    {\sum_\lambda \sum_{\sigma_B}
          |{\cal T}_{\lambda,\sigma_B}|^2}\,,
\end{eqnarray}
where the subscript $\pm 1$ or $0$ denotes the $Z$ helicity.

Combining the $Z$-boson production process (\ref{eq:two_body_decay}) and
the lepton angle distribution (\ref{eq:lepton_angle_distribution}) coherently, $\xi_\ell$ and involving angular distribution linear in
$\cos\theta_\ell$, we obtain the normalized and correlated lepton
angle distribution
\begin{eqnarray}
\frac{1}{\cal C} \frac{{\rm d} {\cal C}}{{\rm d}\cos\theta_\ell}
 = \frac{3}{8}\left[1+\cos^2\theta_\ell
  + P^A_L\, (1-3\cos^2\theta_\ell) + 2\xi_\ell A^A_\pm \cos\theta_\ell\right]\,,
\end{eqnarray}
uniquely determined by the degree of longitudinal $Z$ polarization
$P^A_L$ and the LR asymmetry $A^A_\pm$.

The degree of longitudinal $Z$-polarization and the LR asymmetry from
the two-body decay $A\to BZ$ depend crucially on the $ABZ$ vertex
structure dynamically, restricted by the spin assignments, and on the
masses of the particles $A$ and $B$ kinematically, more specifically
the ratios $z_{A,B}=m_{A,B}/m_Z$. Unless we are concerned about any
overall numerical factors and higher-dimensional couplings such as
those induced from loop corrections, then the $ABZ$ vertex structure
for a given spin assignment is {\it uniquely} determined {\it up to
an overall factor}. For each spin assignment of the particles $A$ and
$B$, the general form of the $ABZ$ vertex structure is given by
\begin{eqnarray}
\mbox{\fbox{\rm S\,\,S}}   &:&  (p+q)^\mu\,,
\label{eq:general_vertex_structure_1}\\
\mbox{\fbox{\rm S\,V}}   &:&  g^{\mu\nu}\,,
\label{eq:general_vertex_structure_2}\\
\mbox{\fbox{\rm V\,S}}   &:& g^{\lambda\mu}\,,
\label{eq:general_vertex_structure_3}\\
\mbox{\fbox{\rm VV}}   &:& g^{\mu\nu} (k-q)^\lambda
               + (q+p)^\mu g^{\nu\lambda}
               - g^{\mu\lambda} (p+k)^\nu\,,
\label{eq:general_vertex_structure_4} \\
\mbox{\fbox{\rm F\,F}}   &:& \gamma^\mu (v+a\gamma_5)\,,
\label{eq:general_vertex_structure_5}
\end{eqnarray}
where $p, q$ and $k$ are the incoming $A$, outgoing $B$ and $Z$
four-momenta, the four-vector indices $\lambda, \nu$ and $\mu$ correspond
to the states $A$ and $B$, and the $Z$ boson, and $v, a$ are the vector
and axial-vector couplings to the $Z$-boson in the FF case.

\begin{figure}[htb]
\centerline{\includegraphics[width=8cm, height=7cm]{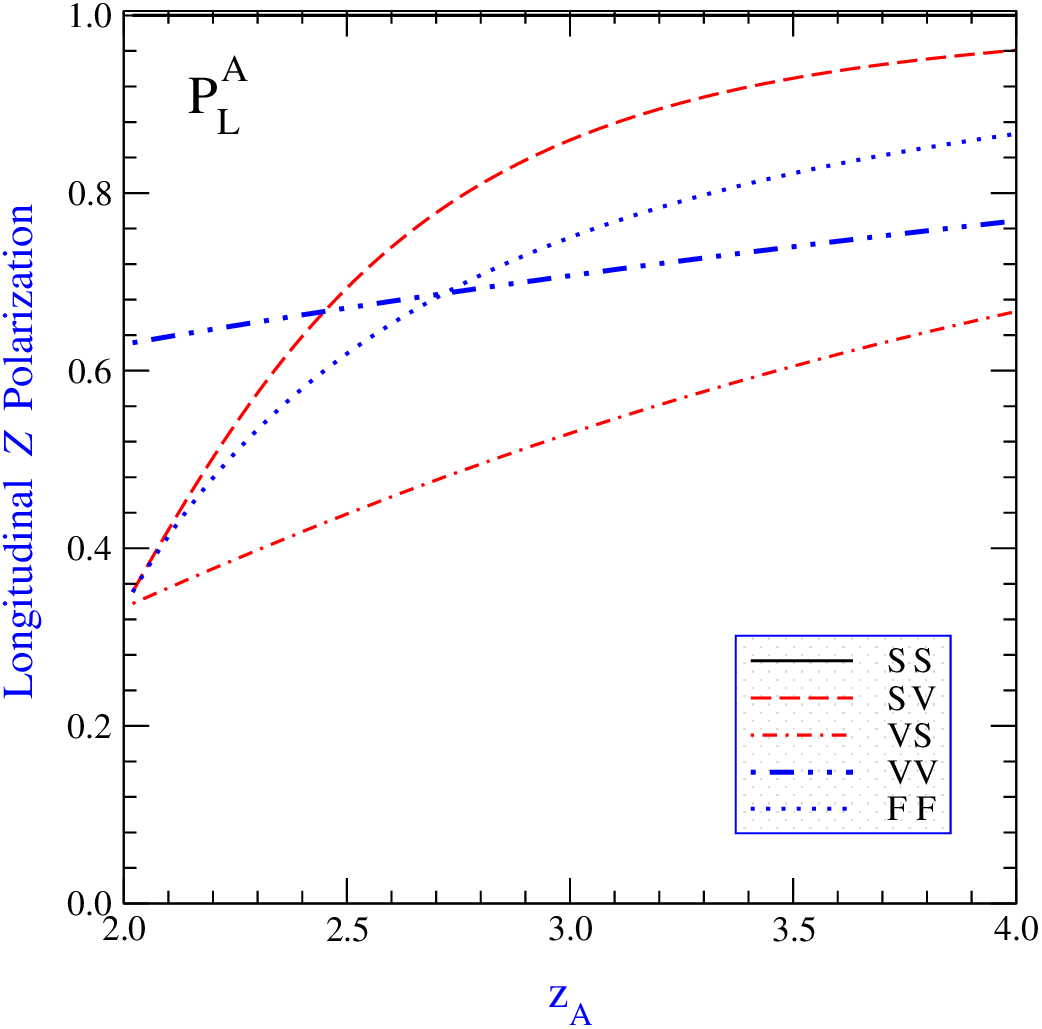}}
\caption{\it The degree of longitudinal $Z$-boson polarization,
          $P^A_L$, as a function of $z_A = m_A/m_Z$ with $z_B=m_B/m_Z=1$
          fixed for each spin assignment of the particles $A$ and $B$
          in the $A$ rest frame. The asymmetries, $\xi_F$ and $A_N$, in
          the FF case are taken to be 0.}
\label{fig:longitudinal_Z_polarization}
\end{figure}

An explicit calculation with the general vertex structures in
Eqs.~(\ref{eq:general_vertex_structure_1}),
(\ref{eq:general_vertex_structure_2}),
(\ref{eq:general_vertex_structure_3}),
(\ref{eq:general_vertex_structure_4}) and
(\ref{eq:general_vertex_structure_5})
yields the analytic expression for
the longitudinal $Z$ polarization for each spin assignment of the
states $A$ and $B$ as
\begin{eqnarray}
&& P^A_L[{\rm S\,\,S}] = 1\,,
   \label{eq:PL_SS} \\
&& P^A_L[{\rm S\,V}]   = (z^2_A-z^2_B-1)^2/
                       \left[(z^2_A-z^2_B-1)^2+ 8z^2_B\right]\,,
   \label{eq:PL_SV} \\
&& P^A_L[{\rm V\,S}]   = (z^2_A-z^2_B+1)^2/
                       \left[(z^2_A-z^2_B+1)^2+8 z^2_A\right]\,,
   \label{eq:PL_VS}\\
&& P^A_L[{\rm VV}] = \left[9 z^2_A z^2_B+2(z^2_A+z^2_B)+1\right]/
                   \left[9 z^2_A z^2_B+10(z^2_A+z^2_B)+1\right]\,,
   \label{eq:PL_VV}\\
&& P^A_L[{\rm F\,F}] = (e_F + \beta^2)/(3\, e_F+ \beta^2)\,,
   \label{eq:PL_FF}
\end{eqnarray}
and the LR asymmetry, non-vanishing only for the FF assignment,
as
\begin{eqnarray}
A^A_\pm [FF] = \xi_F \frac{2\beta}{3\, e_F +\beta^2}\quad \mbox{with}
\quad \xi_F = \frac{2va}{v^2+a^2}\,,
\label{eq:A_LR_FF}
\end{eqnarray}
where, for notational convenience, we have used the abbreviations
\begin{eqnarray}
\beta &=& \lambda^{1/2}(z^2_A, z^2_B, 1)\,,\\
e_F &=& z^2_A+z^2_B-1- 2z_A z_B A_N\,,
\end{eqnarray}
with $z_{A,B}=m_{A,B}/m_Z$ and the asymmetry $A_N= (v^2-a^2)/(v^2+a^2)$
defined in terms of the vector and axial-vector couplings $v$ and $a$.
In order for the two-body decay $A\to BZ$ to be kinematically allowed,
the inequality $z_A \geq z_B + 1$ should be satisfied.

One aspect unique to the SS spin assignment is that the $Z$-boson is
completely longitudinally polarized independently of the masses $m_A$
and $m_B$. Another distinctive feature shared by all the spin assignments
is that the $Z$ boson becomes longitudinally polarized as
$z_A\to \infty$, {\it cf.} Figure~\ref{fig:longitudinal_Z_polarization}.

\section{$Z$ Polarization in the Laboratory Frame}

Although the $Z$ boson itself is fully reconstructed,  the rest of the
event, in particular, the rest frame of the decaying particle $A$, may
not be reconstructed. In this situation, a natural axis for the
$Z$-polarization is the $Z$-boson's momentum direction in the laboratory
frame, which will be called the detection axis.
The most natural experimental observable for $Z$ decays is then the polar-angle $\theta^\prime$ distribution of the charged lepton
$\ell^-$ with respect to the $Z$-boson direction of motion in the
$Z$-boson rest frame.

The degree of longitudinal $Z$ polarization $P^D_L$ and the LR asymmetry
$A^D_\pm$ determined along the detection axis is related to the degree of
longitudinal $Z$ polarization $P^A_L$ and the LR asymmetry $A^A_\pm$
computed in the rest frame of the decaying particle $A$ by a Wigner
rotation connecting the two spin bases~\cite{Wigner_rotation}.
The detected polarization $P^D_L$ and the LR asymmetry $A^D_\pm$ are
given by
\begin{eqnarray}
P^D_L &=& \cos^2\!\omega\, P^A_L
        + \frac{1}{2}\sin^2\!\omega\, (1-P^A_L)\,, \\
A^D_\pm &=& \cos\!\omega\, A^A_\pm\,,
\end{eqnarray}
in terms of the computed polarization $P^A_L$ and LR asymmetry $A^A_\pm$,
respectively, and the Wigner angle $\omega$ between the production and
detection axes in the $Z$ rest frame. This angle is determined by the
composition of boosts $\Lambda_{ZL}$ from the $Z$ rest frame to the
laboratory frame, followed by $\Lambda_{LP}$ from the laboratory frame
to the $A$ rest frame, and finally $\Lambda_{PZ}$ from the $A$ rest frame
to the $Z$ rest frame,
\begin{eqnarray}
  {\cal M}(\Lambda_{ZL}){\cal M}(\Lambda_{LP}){\cal M}(\Lambda_{PZ})
= {\cal R}(\omega)\,,
\end{eqnarray}
where ${\cal M}(\Lambda)$ and ${\cal R}(\theta)$ are the representation
matrices for the Lorentz transformations. Explicitly, the Wigner angle
$\omega$ can be extracted from the expression
\begin{eqnarray}
\tan\omega = \frac{m_Z\beta_A\sin\Theta}{p_Z + \beta_A E_Z \cos\Theta}
           = \frac{\beta_A\sin\Theta}{\gamma_Z
                                      (\beta_Z+\beta_A \cos\Theta)}\,,
\end{eqnarray}
where $E_Z=\gamma_Z\, m_Z$ and $p_Z=\gamma_Z\beta_Z\, m_Z$ are the energy
and absolute momentum of the $Z$ in the $A$ rest frame, $\Theta$ is
the polar angle of the $Z$ boson in the $A$ rest frame and $\beta_A$ is
the speed of the particle $A$ in the laboratory frame.

The energy and absolute momentum of the $Z$ boson are fixed with the masses
of the particles in the decay $A\to BZ$ as
\begin{eqnarray}
E_Z &=& \frac{m_Z}{2}\, \frac{(z^2_A-z^2_B+1)}{z_A}\,,
\label{eq:Z_energy} \\
p_Z &=& \frac{m_Z}{2}\, \lambda^{1/2}(z^2_A, z^2_B, 1)\,,
\label{eq:Z_momentum}
\end{eqnarray}
The phase space integration over $\cos\Theta$ can be carried out as the
matrix elements are independent of $\cos\Theta$ after averaging over
the $A$ and $B$ spins.

The only nontrivial complication arises from the boost spectrum
${\cal D}(\beta_A)$ between the $A$ rest frame and the laboratory frame
which depends on the $A$ production mechanism and the parton distribution
functions. In principle the (normalized) distribution ${\cal D}$ can be
computed for a given model. If so, the averages of the degree of observed
longitudinal polarization and the observed LR asymmetry are given by
\begin{eqnarray}
\langle\langle P^D_L \rangle\rangle
=  P^A_L -\langle\langle\, \sin^2\!\omega \rangle\rangle\,
          \frac{1}{2} \, \left(3P^A_L-1\right)
   \quad\mbox{and}\quad
   \langle\langle A^D_\pm \rangle\rangle
= \langle\langle\, \cos \omega \rangle\rangle\,
   A^A_\pm\,,
\end{eqnarray}
with the double-integration expressions defined by
\begin{eqnarray}
   \langle\langle\, \sin^2\!\omega/\cos \omega \rangle\rangle
=  \int^1_0 d\beta_A\, {\cal D}(\beta_A)\, \frac{1}{2} \int^{1}_{-1} d\cos\Theta\,
   \, \sin^2\!\omega/\cos \omega (\Theta, \beta_A)\,,
\end{eqnarray}
that allow fully detailed quantitative predictions for the spin
assignments of the particles $A$ and $B$.

There are two extreme kinematic limits for which we do not have to rely
on any detailed information on the boost distributions in practice.
Firstly, if the particle $A$ is produced near threshold with
$\beta_A\rightarrow 0$, then $\cos\omega \rightarrow 1$ rendering
the difference between the production and detection axes negligible.
Secondly, if the mass splitting, $m_A-m_B$, of the particles $A$ and $B$
is much larger than $m_Z$. In this case the $Z$ boson is highly boosted
with $E_Z$ and $p_Z$ much larger than $m_Z$ even in the $A$ rest frame
except for the far backward region with $\Theta$ very close to $\pi$.

In order to analyze the boost dependence quantitatively, we consider
the averages of $\sin^2\!\omega$ and $\cos\omega$ over the angle $\Theta$,
$\langle \sin^2\!\omega/\cos\omega\rangle=\frac{1}{2}\int d\cos\Theta\,
\sin^2\!\omega/\cos\omega$, of which the analytic forms are
\begin{eqnarray}
  \langle \sin^2\!\omega \rangle (\beta_A)
& =&  \frac{1}{\gamma^2_Z\beta^3_Z\beta_A}
  \bigg[\frac{1}{\beta_Z\beta_A}
        \ln\left|\frac{\beta_Z+\beta_A}{\beta_Z-\beta_A}\right|
        \nonumber\\
  && \mbox{ } \hskip 1.5cm
       -\frac{(\gamma^2_Z+\gamma^2_A)}{2\gamma_Z\beta_Z\gamma_A\beta_A}
        \ln\left|\frac{\gamma_Z\beta_Z+\gamma_A\beta_A}{
                       \gamma_Z\beta_Z-\gamma_A\beta_A}\right|
       -1
  \bigg]\,, \\
 \langle \cos\omega\rangle (\beta_A)
& = & \frac{1}{2\beta^2_Z\beta_A}
  \bigg[\beta_Z+\beta_A-|\beta_Z-\beta_A|
        \nonumber\\
  && \mbox{ }\hskip 1.5cm
       -\frac{1}{\gamma^2_Z}
        \ln\left(\frac{1+\beta_Z\beta_A+\beta_Z+\beta_A}{
                       1-\beta_Z\beta_A+|\beta_Z-\beta_A|}\right)
  \bigg]\,,
\end{eqnarray}
as a function of $\beta_A$, respectively, satisfying the boundary
condition that
$\langle \sin^2\!\omega\rangle (0)= 0$ and
$\langle \cos\omega\rangle (0)=1$.
Here, the $Z$-boson boost factors, $\gamma_Z= E_Z/m_Z$ and
$\beta_Z=p_Z/E_Z$,
with $E_Z$ and $p_Z$ in Eqs.~(\ref{eq:Z_energy}) and
({\ref{eq:Z_momentum}) fixed with the masses $m_A$ and $m_B$
in the decay. As $\beta_A\rightarrow 1$,
the two average functions approach asymptotically to
\begin{eqnarray}
  \langle \sin^2\!\omega \rangle (1)
&=& \frac{1}{\gamma^2_Z\beta^2_Z}
  \left[ \frac{1}{\beta_Z}\ln \left(\frac{1+\beta_Z}{1-\beta_Z}\right) -2\right] \ \ \ \
  \rightarrow \ \  \frac{2}{3}\ \ \ \mbox{as}\ \ \ \beta_Z \to 0\,, \\
 \langle \cos\omega\rangle (1)
&=& \frac{1}{\gamma^2_Z\beta^2_Z}
  \left[ \beta_Z-\frac{1}{2}
        \ln\left(\frac{1+\beta_Z}{1-\beta_Z}\right)\right] \ \ \ \
  \rightarrow \  \ 0\ \ \ \mbox{as}\ \ \ \beta_Z \to 0\,.
\end{eqnarray}
As a numerical illustration, the dependence of the averages
$\langle \sin^2\!\omega\rangle$ and
$\langle \cos\omega\rangle$ on the boost $\beta_A$ is shown in
Figure~\ref{fig:polarization_reduction_factor} for three
characteristic sets of new particle masses,
$(m_A, m_B) = (0.5,0.4), (0.6,0.4)$ and $(0.9, 0.3)\, [{\rm TeV}]$
with $m_Z=91$\, GeV.

\vskip 0.5cm
\begin{figure}[thb]
\centerline{
\includegraphics[width=5.5cm, height=6.5cm]{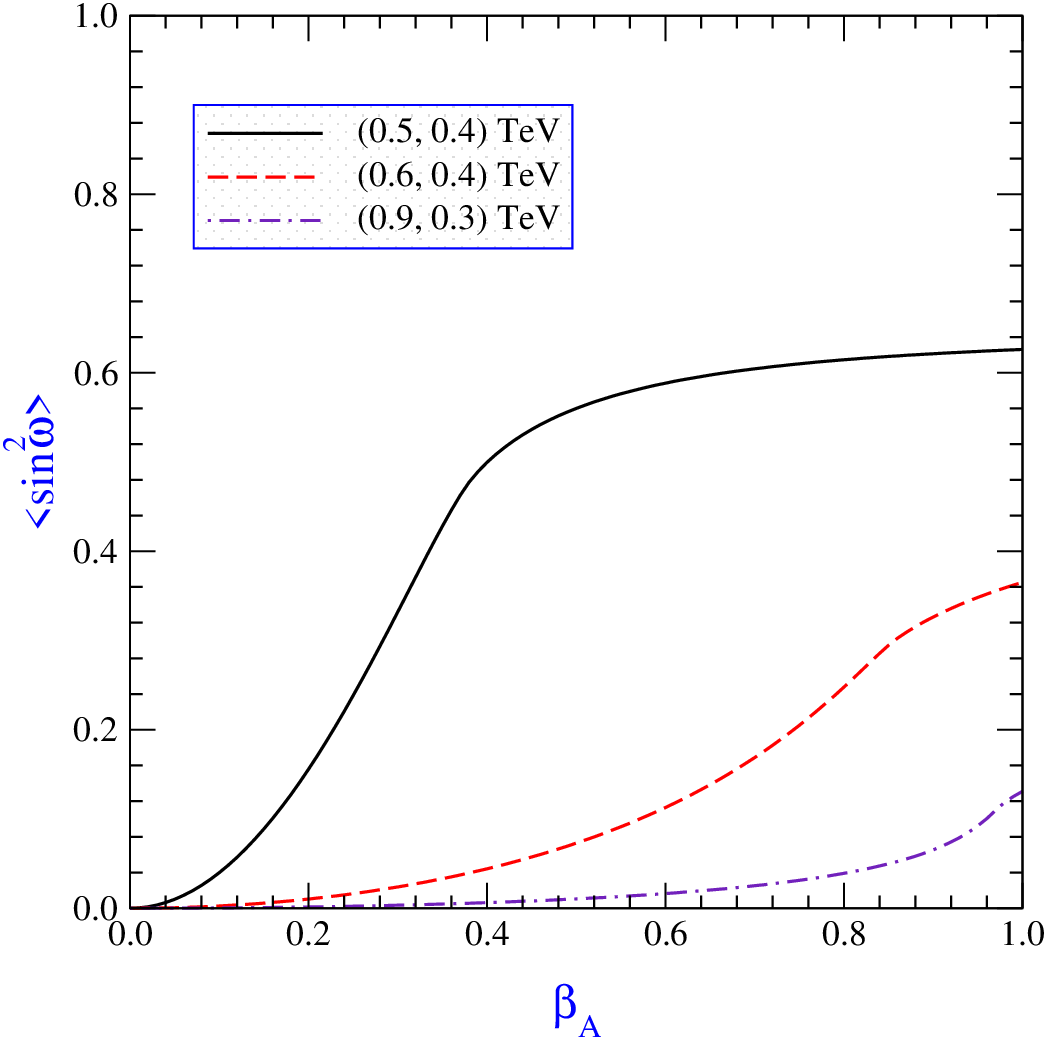}
\hskip 1.0cm
\includegraphics[width=5.5cm, height=6.5cm]{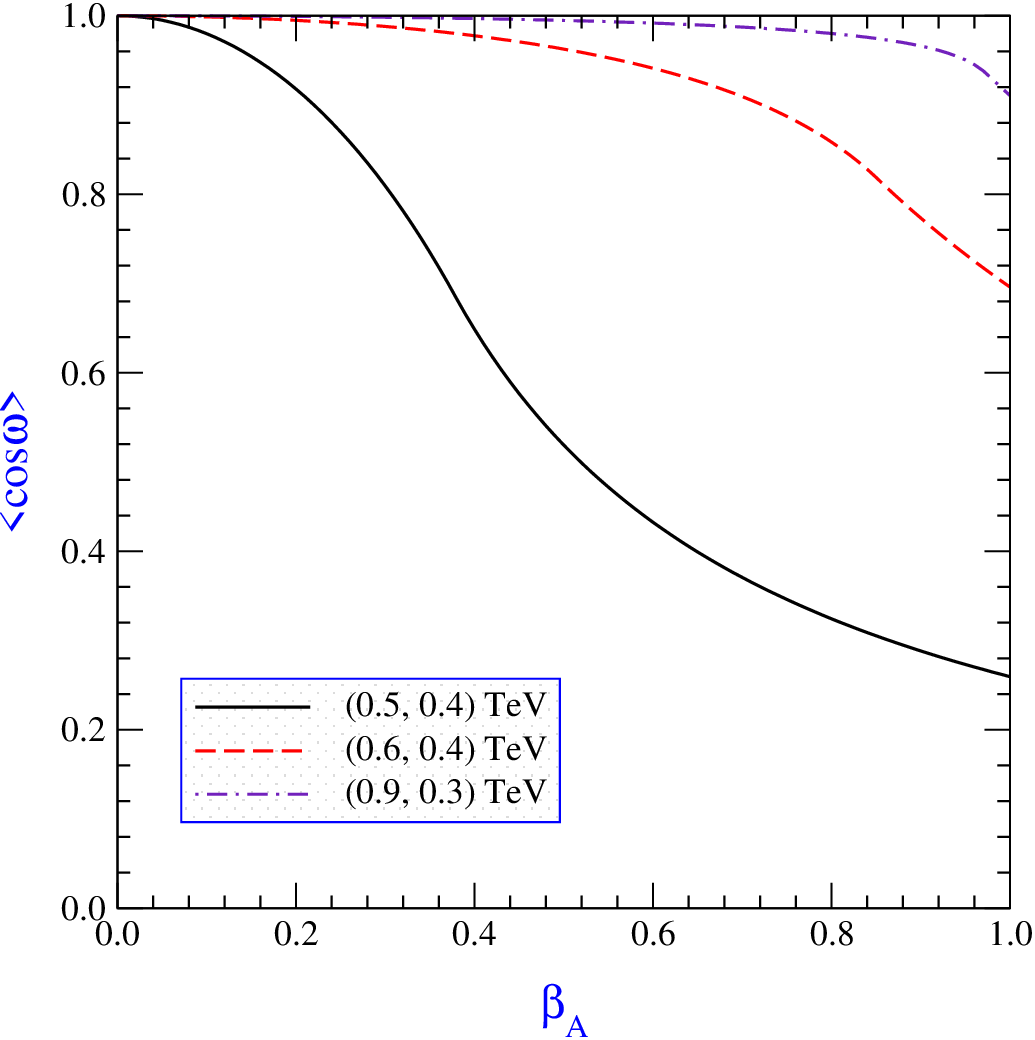}}
\caption{\it The dependence of the averages
         $\langle \sin^2\!\omega\rangle$ (left frame)
         and $\langle \cos\omega\rangle$ (right frame)
         as a polarization reduction factor on the boost $\beta_A$.
         For the purpose of illustration, three sets of new particle
         masses, $(m_A, m_B) = (0.5,0.4), (0.6,0.4)$
         and $(0.9, 0.3)\, [{\rm TeV}]$ with $m_Z=91$\, GeV are chosen.
        }
\label{fig:polarization_reduction_factor}
\end{figure}

\section{Conclusions}

We have computed the expected $Z$ polarization
for various spin assignments from a well-motivated class of
models beyond the SM. In addition, we have provided a detailed analytic
description about how to determine the degree of $Z$-boson polarization
even if the decaying particle is not at rest.
As demonstrated numerically, $Z$ bosons produced in any two-body decay
$A\to B Z$ involving two new particles $A$ and $B$ can provide us with
an important and powerful handle for determining the spins of the new
particles, unless $Z$ is not at rest nor boosted with extremely high
energies.

\section*{Acknowledgments}

The author thanks A. Freitas and Y.G. Kim for their collaboration on
the work at its early stage. The work was supported in part by the
CERN-Korea theory collaboration program (NRF-2012K1A3A2A01051781) and
in part by Basic Science Research Program through the National
Research Foundation (NRF) funded by the Ministry of Education,
Science and Technology (NRF-2016R1D1A3B01010529).



\begin{thebibliography}{99}

\bibitem{Weinberg:1976}
  S. Weinberg,
  Phys. Rev. D {\bf 13}, 974 (1976).

\bibitem{Weinberg:1979bn}
  S. Weinberg,
  Phys. Rev. D {\bf 19}, 1277 (1979).

\bibitem{Susskind:1979}
  L. Susskind,
  Phys. Rev. D {\bf 20}, 2619 (1979).

\bibitem{tHooft:1980}
  G. 't Hooft, in {\it Recent developments in gauge theories :
  Proceedings of the NATO Advanced Summer Institute, Cargese 1979},
  (Plenum, New York 1980).

\bibitem{Nilles}
  H. P. Nilles,
  Phys. Rept. {\bf 110}, 1 (1984).

\bibitem{Haber_Kane}
  H. E. Haber and G. L. Kane,
  Phys. Rept. {\bf 117}, 75 (1985).

\bibitem{Chung:2003fi}
  D. J. H. Chung {\it et al,},
  Phys. Rept. {\bf 407}, 1 (2005)

\bibitem{Drees}
   M. Drees, R. Godbole and P. Roy,
   {\it Theory and Phenomenology of Sparticles: An Account of Four-dimensional $N=1$ Supersymmetry in High Energy Physics},
   (World Scientific, Hackensack, 2004).

\bibitem{Binetruy}
  P. Binetruy, {\it Supersymmetry: Theory, experiment and cosmology},
  (Oxford University Press, Oxford, 2006).

\bibitem{Wess_Bagger}
  J. Wess and J. Bagger, {\it Supersymmetry and supergravity},
  (Princeton University Press, Princeton, 1992).

\bibitem{Appelquist:2000nn}
  T. Appelquist, H. C. Cheng and B. A. Dobrescu,
  Phys. Rev. D {\bf 64}, 035002 (2001)
  [arXiv:hep-ph/0012100].

\bibitem{Wang:2006hk}
  A. J. Barr,
  Phys. Lett. B {\bf 596}, 205 (2004)
  [arXiv:hep-ph/0405052].

\bibitem{Smillie:2005ar}
  J. M. Smillie and B. R. Webber,
  JHEP {\bf 0510}, 069 (2005)
  [arXiv:hep-ph/0507170].

\bibitem{Wang_Yavin}
  L. T. Wang and I. Yavin,
  JHEP {\bf 0704}, 032 (2007)
  [arXiv:hep-ph/0605296].

\bibitem{Wang:2008sw}
  L. T. Wang and I. Yavin,
  arXiv:0802.2726 [hep-ph].

\bibitem{Choi:2003fs}
  S. Y. Choi and Y. G. Kim,
  Phys. Rev. D {\bf 69}, 015011 (2004)
  [arXiv:hep-ph/0311037].

\bibitem{Kim:2007zzm}
  Y. G. Kim,
  Phys. Rev. D {\bf 76}, 077702 (2007).

\bibitem{Jacob_Wick}
  M. Jacob and G. C. Wick, Annals Phys. {\bf 7}, 404 (1959);
  {\it ibid.} {\bf 281}, 774 (2000).

\bibitem{Wigner_rotation}
   See, e.g., AD. Martin and T.D. Spearman,
   {\it Elementary particle theory},
   (North-Holland, Amsterdam, 1970).

\end{thebibliography}
\end{document}